\documentclass[a4paper,fleqn]{article}
\usepackage{graphicx}
\usepackage[text={17.8cm,22.0cm},centering]{geometry}
\usepackage[numbers]{natbib}

\usepackage{amsmath}
\usepackage{amssymb}
\usepackage{amsthm}
\usepackage{enumerate}
\usepackage{epsfig}
\usepackage{epstopdf}
\usepackage{amstext}
\usepackage{textcomp}
\usepackage{amsfonts}
\usepackage{color}
\usepackage{float}
\usepackage{tabularx}
\usepackage{tabulary}
\usepackage{graphicx}

\newtheorem{Theorem}{Theorem}

\newtheorem{Assumption}{Assumption}%[section]
\newtheorem{Definition}{Definition}%[section]

\begin{document}

\begin{center}
	{\bf Coupled Fixed Points for Hardy--Rogers Type of Maps and Their Applications in the Investigations of Market Equilibrium in Duopoly Markets for Non--Differentiable, Nonlinear Response Functions}
\end{center}

\begin{center}
	\begin{tabular}{c}
	Stanimir Kabaivanov\\
	Faculty of Economics and Social Sciences,\\
	University of Plovdiv Paisii Hilendarski,\\
	4000 Plovdiv, Bulgaria\\
	stanimir.kabaivanov@uni-plovdiv.bg\\
	orcid:0000-0002-8686-8112
	\end{tabular}
\end{center}

\begin{center}
	\begin{tabular}{c}
		Vasil Zhelinski\\
		Faculty of Mathematics and Informatics,\\
		University of Plovdiv Paisii Hilendarski,\\
		4000 Plovdiv, Bulgaria\\
		vasilzhelinski@abv.bg
		\end{tabular}
\end{center}

\begin{center}
	\begin{tabular}{c}
		Boyan Zlatanov\\
		Faculty of Mathematics and Informatics,\\
		University of Plovdiv Paisii Hilendarski,\\
		4000 Plovdiv, Bulgaria\\
		bobbyz@uni-plovdiv.bg\\
		orcid:0000-0002-5857-9372
	\end{tabular}
\end{center}

{\sc abstract:}
	In this paper we generalize Hardy--Rogers maps in the context of coupled fixed points. We generalizes with the help of the obtained main theorem 
	some known results about existence and uniqueness of market equilibrium in duopoly markets. We investigate some recent results about market equilibrium 
	in duopoly markets with the help of the main theorem and we enrich them. We define a generalized response function including production and surpluses. Finally we 
	illustrate a possible application of the main result in the investigation of market equilibrium, when the pay off functions are non differentiable.
	
{\bf keywords:}
	duopoly equilibrium; response functions; fixed points; contraction map; coupled fixed points
	
\section{Introduction}

The investigation of coupled fixed points started in 1987 from the article \cite{GL}. 
The first result \cite{GL} deals with maps with the mixed monotone property in complete partially ordered metric spaces.
There is a great number of conteporary research papers in the theory of coupled fixed points with the mixed monotone property \cite{Raden}, without the mixed monotone property \cite{DKR}, in ordered probabilistic metric spaces \cite{THO}, in modular metric spaces \cite{ShJ}, in a metric spaces endowed with a graph \cite{BMShA}, fuzzy cone metric spaces \cite{WRJGA}, in $b$--metric spaces \cite{KNM}, for multi-valued maps \cite{GhM} .
The idea of coupled fixed points was generalized for coupled best proximity points \cite{SKumam}. There are applications of coupled fixed points in different fields of mathematics:
impulsive differential equations \cite{BMShA}, integral equations \cite{KChDS}, ordinary differential equations \cite{KhTa}, periodic boundary value problem \cite{FaNi}, fractional equations \cite{NaIb}, nonlinear matrix equations \cite{RaEl} in other sciences: economics \cite{DKRZ,KaZl}, aquatic ecosystems \cite{GHNRZ}, dynamic programming \cite{ChMK}, which are just the most recent investigation
that are dealing with coupled fixed points.

Important results about the connection between coupled fixed points and fixed points is obtained in \cite{Petrusel}. Following the suggested in \cite{Petrusel} technique we present a generalization of Hardy--Rogers maps in the coupled fixed point theory. 

The analysis of market equilibrium in duopoly markets was pioneered by \cite{Cournot} in 1897. Due to its practical importance it is still a metter of high interest and rapid development nowadays \cite{UdMiRa,Gut}. 
The classical approach is to maximize the pay off function of the two participants in the market \cite{OS,BCKS,MS} with the justification that rational players will always try to gain maximum profitality in the real world. However the contemporary business conditions offer more complex goals, including multi-period targets, market share gains to name a few. Therefore, a different approach has been suggested in by considering the response functions of both players, rather than maximizing their pay off functions. While this may look unsual at first glance, there are a large number of solid arguments that immediate (or current period) profits may not provide a good guidance in understanding company actions and describing duopoly markets.
First of all, the complexity and size of business organizations vary, which means that the time and way of execution of high level goals will also do so. As a result, the goal for achieving higher profits may actually be implemented by following different sub-goals (like for example ensuring higher engagement of workforce), providing for higher core growth rate or agressing international expansion. It is often the case, that more a combination of sub-goals is involved in the company strategic plans. In case of different master plans, business entities may react differently on changes in economic conditions. While for an outside observer these could look like non-optimal responses, it may well be that they are aligned with the long-term strategy, rather than current period profitabiilty.
Non-homogeneous products are also important in studying company behavior in duopoly markets. While one can argue that there are a lot of situations where the products may be virtually identical (like for example raw materials), a sound market model should also be able to describe cases where output characteritics differ. As a result, companies may stess on different approaches to gain competitive advantage - in particular those that are not based on price changes.
Time for change, or time required to implement a high level decision may be different, depending on company organizational structure, size and product diversity. There are certain decisions (like for example price change) that may be implemented right away and bring also quicker results. Others however require more time, or may be even considered unfeasiable due to technological or scale concerns (for exapmle change in production levels may not be possible at any required number due to batch processing, minimum number of outputs or technological concerns that require a certain number of products to be manufactured at once). As a result the outputs are not perfectly divisable and that imposes an additional restriction on how companies react and behave in duopoly markets. 

We have tried to generalize these ideas from by showing that the maximization problem may lack the second order conditions, do to a nondifferentiabilty of the pay off functions. 

\section{Preliminaries}

\begin{Definition}(\cite{Hardy-Rogers})\label{def-HR}
	Let $(X,\rho)$ be a metric space. A map $T:X\to X$ is a Hardy--Rogers map if there are nonnegative constants $a_i$, $i=1,2,3,4,5$, satisfying $\sum_{i=1}^5 a_i<1$, so that for any $x,y\in X$ holds the inequality
	\begin{equation}\label{eq:1}
		\rho(Tx,Ty)\leq a_1\rho(x,y)+a_2\rho(x,Tx)+a_3\rho(y,Ty)+a_4\rho(x,Ty)+a_5\rho(y,Tx).
	\end{equation}
\end{Definition}

As pointed in \cite{Hardy-Rogers} from the symmetry of the metric function $\rho(\cdot,\cdot)$ it follow if (\ref{eq:1}) holds true
then it will hold the inequality 
$\rho(Tx,Ty)\leq a_1\rho(x,y)+\frac{a_2+a_3}{2}(\rho(x,Tx)+\rho(y,Ty))+\frac{a_4+a_5}{2}(\rho(x,Ty)+a_5\rho(y,Tx))$.
Therefore, without loss of generality we can consider maps that
satisfy the inequality
$$
\rho(Tx,Ty)\leq k_1\rho(x,y)+k_2(\rho(x,Tx)+\rho(y,Ty))+k_3(\rho(x,Ty)+\rho(y,Tx)),
$$
so that $k_1+2k_2+2k_3<1$.

If $k_2=k_3=0$ we get a Banach contraction map, if $k_1=k_3=0$ we get a Kannan contraction map \cite{Kannan} and if $k_1=k_2=0$ we get a Chatterjea contraction map \cite{Chatterjea}.

\begin{Theorem}(\cite{Hardy-Rogers})\label{th-HR}
	If $(X,\rho)$ is a complete metric space and $T:X\to X$ be a Hardy--Rogers map, then 
	\begin{enumerate}\itemsep=-4pt
		\item there is a unique fixed point $\xi\in X$ of $T$ and moreover for any initial guess $x_0\in x$ the iterated sequence $x_n=Tx_{n-1}$, for $n=1,2,\dots$ converges to the fixed point $x$
		\item there holds the a priori error estimates
		$\rho(\xi,x_n)\leq \frac{k^n}{1-k}\rho(x_0,x_1)$
		\item there holds a posteriori error estimate
		$\rho(\xi,x_n)\leq \frac{k}{1-k}\rho(x_{n-1},x_n)$
		\item the rate of convergence 
		$\rho(\xi,x_n)\leq k\rho(\xi,x_{n-1})$,
	\end{enumerate}
	where $k=\frac{k_1+k_2+k_3}{1-k_2-k_3}$.
\end{Theorem}

\begin{Definition}[\cite{GL}]\label{defGL}
	Let $A$ be nonempty subset of a metric space $(X,\rho)$, $F:A\times A\to A$. An ordered pair $(\xi,\eta)\in A\times A$ is said to be a coupled fixed point of $F$ in $A$ if $\xi=F(\xi,\eta)$ and $\eta=F(\eta,\xi)$.
\end{Definition}

In order to apply the technique of coupled fixed points, in invetigation of market equilibrium, a generalization of the mentioned above notions was presented in \cite{DKRZ}.
As far as in duopoly markets naturally each of the players' has a different response reaction to its rival production and the market one.
Thus two response functions $F_i$ for $i=1,2$ were considered in \cite{DKRZ}, such that  $F_i:X_1\times X_2\to X_i$, $i=1,2$,
where $X_i$ is the production set of player $i=1,2$ and the coupled fixed points were defined by $x=F_1(x,y)$ and $y=F_2(x,y)$.
Whenever $X_1=X_2=A$ and $F_2(x,y)=F_1(y,x)$ we get notion of coupled fixed points from Definition \ref{defGL}.

\subsection{Main Result}

We will generalize the notions from \cite{DKRZ}, by considering two different metric spaces $(Z_i,d_i)$, $i=1,2$.

\begin{Definition}\label{new-fixed-point}
	Let $X_1$, $X_2$ be nonempty subsets of the metric spaces $(Z_1,d_1)$ and $(Z_1,d_2)$, respectively, 
	$F_i:X_1\times X_2\to X_i$ for $i=1,2$. 
	An ordered pair $(\xi,\eta)\in X_1\times X_2$ is called a coupled fixed point of $(F_1,F_2)$ if
	$\xi=F_1(\xi,\eta)$ and $\eta=F_2(\xi,\eta)$.
\end{Definition}

\begin{Definition}\label{iterated_sequence}
	Let $X_1$, $X_2$ be nonempty subsets of the metric spaces $(Z_1,d_1)$ and $(Z_2,d_2)$, respectively, 
	$F_i:X_1\times X_2\to X_i$ for $i=1,2$.
	For any pair $(x,y)\in X_1\times X_2$ we define the sequences $\{x_n\}_{n=0}^\infty$ and $\{y_n\}_{n=0}^\infty$ by
	$x_0=x$, $y_0=y$ and $x_{n+1}=F_1(x_{n},y_{n})$, $y_{n+1}=F_2(x_{n},y_{n})$ for all $n\geq 0$.
\end{Definition}

Everywhere, when considering the sequences $\{x_n\}_{n=0}^\infty$ and $\{y_n\}_{n=0}^\infty$ 
we will assume that they are the sequences defined in Definition \ref{iterated_sequence}.

\begin{Theorem}\label{th-HR-new}
	Let $(X_1,d_1)$ and $(X_2,d_2)$ be two complete metric spaces. Let there are two maps
	$F_i:X_1\times X_2\to X_i$, for $i=1,2$ and let there are non--negative constants $k_i$ for $i=1,2,3$, so that $k_1+2k_2+2k_3<1$ and the ordered pair of maps
	$(F_1,F_2)$ satisfies the inequality
	\begin{equation}\label{eq-HR}
		\begin{array}{lll}
			\sum_{i=1}^2d_i(F_i(x,y),F_i(u,v))&\leq& k_1(d_1(x,u)+d_2(y,v))\\
			&&+
			k_2(d_1(x,F_1(x,y))+d_2(y,F_2(x,y))+d_1(u,F_1(u,v))+d_2(v,F_2(u,v)))\\
			&&+k_3(d_1(x,F_1(u,v))+d_2(y,F_2(u,v))+d_1(u,F_1(x,y))+d_2(v,F_2(x,y)))
		\end{array}
	\end{equation}
	for any $(x,y),(u,v)\in X_1\times X_2$.
	Then
	
	\begin{enumerate}\itemsep=-4pt
		\item there is a unique coupled fixed point $(\xi,\eta)\in X_1\times X-2$ of $(F_1,F_2)$ and moreover for any initial guess $(x_0,y_0)\in x$ the iterated sequences $x_n=F_1(x_{n-1},y_{n-1})$ and $y_n=F_2(x_{n-1},y_{n-1})$, for $n=1,2,\dots$ converge to the coupled fixed point $(\xi,\eta)$
		\item there holds the a priori error estimates
		$d_1(\xi,x_n)+d_2(\eta,y_n)\leq \frac{k^n}{1-k}(d_1(x_0,x_1)+d_2(y_0,y_1))$
		\item there a posteriori error estimate
		$d_1(\xi,x_n)+d_2(\eta,y_n)\leq  \frac{k}{1-k}(d_1(x_n,x_{n-1})+d_2(y_n,y_{n-1}))$
		\item the rate of convergence 
		$d_1(\xi,x_n)+d_2(\eta,y_n)\leq k(d_1(\xi,x_{n-1})+d_2(\eta,y_{n-1}))$
	\end{enumerate}
	where $k=\frac{k_1+k_2+k_3}{1-k_2-k_3}$.
	
	If in addition $X_1,X_2\subseteq X$, where $(X,d)$ is a complete metric space and $F_2(x,y)=F_1(y,x)$, then the coupled fixed point $(\xi,\eta)$ satisfies $\xi=\eta$.
\end{Theorem}

{\bf Remark:} By $F_2(x,y)=F_1(y,x)$, actually we assume that $F_i$ is defined on the set $(X_1\cup X_2)\times (X_1\cup X_2)$, but
it is possible for example $F_1(x_1,x_2)$, where $x_1,x_2\in X_1$ not to belong to the set $X_1$ or not to hold inequality (\ref{eq-HR}).

\proof
Let us consider the product space $(X_1\times X_2,\rho)$, endowed with the metric
$\rho(\cdot,\cdot)=d_1(\cdot,\cdot)+d_2(\cdot,\cdot)$. From the assumption that 
$(X_i,d_i)$ be complete metric spaces it follows that $(X_1\times X_2,\rho)$ is a complete metric space too.

Following \cite{Petrusel} let us define a map $G:X\times Y\rightarrow X\times Y$ by $G(x,y)=(F_1(x,y),F_2(x,y))$.
Then inequality (\ref{eq-HR}) is equivalent to
\begin{equation}\label{eq-HR-new}
	\begin{array}{lll}
		\rho(G(x,y),G(u,v))&\leq& k_1\rho((x,y),(u,v))+
		k_2(\rho((x,y),G(x,y))+\rho((u,v),G(u,v))\\
		&&+k_3(\rho((x,y),G(u,v))+\rho((u,v),G(x,y))
	\end{array}
\end{equation}
and therefore the map $G:X_1\times X_2\to X_1\times X_2$ is a Hardy--Rogers map in the complete metric space $(X_1\times X_2,\rho)$. Consequently we can apply Theorem \ref{th-HR}
and we will get that there is a unique $(\xi,\eta)\in X_1\times X_2$, such that
$(\xi,\eta)=G(\xi,\eta)=(F_1(\xi,\eta),F_2(\xi,\eta))$, i.e $\xi=F_1(\xi,\eta)$ and
$\eta=F_2(\xi,\eta)$. The error estimates followed directly form the definition of the metric $\rho$ and Theorem \ref{th-HR}.

If $X_1,X_2\subseteq X$, $F_2(x,y)=F_1(y,x)$ and $d_1=d_2=d$, then for the fixed point $(x,y)$ of the map $G$, by using of (\ref{eq-HR}) with $u=y$ and $v=x$, we get
\begin{equation}\label{x3}
	\begin{array}{lll}
		2d(x,y)&=&2d(F_1(x,y),F_2(x,y))=d(F_1(x,y),F_1(y,x))+d(F_2(x,y),F_2(y,x))=\rho(G(x,y),G(y,x))\\
		&\leq& 2k_1 d(x,y)+k_2(d(x,F_1(x,y))+d(y,F_2(x,y))+d(y,F_1(y,x))+d(x,F_2(y,x)))\\
		&&+k_3(d(x,F_1(y,x))+d(y,F_2(y,x))+d(y,F_1(x,y))+d(x,F_2(x,y)))\\
		&=& 2k_1 d(x,y)+k_2(d(x,x)+d(y,y)+d(y,y)+d(x,x))+k_3(d(x,y)+d(y,x)+d(y,x)+d(x,y))\\
		&=& 2k_1\rho(x,y)+2k_3\rho(x,y)<2d(x,y),
	\end{array}
\end{equation} which is a contradiction and therefore $x=y$.

\section{\normalsize Applications of Theorem \ref{th-HR}}

We will present some corollaries of Theorem \ref{th-HR}.

\subsection{\normalsize Generalization of some known results about coupled fixed points and corollaries of Theorem \ref{th-HR-new}}
Let us recall the main result from \cite{DKRZ}.

\begin{Theorem}\label{th:3456}
	Let $X_1$, $X_2$ be nonempty and closed subsets of a complete metric space $(X,d)$. Let there exist a closed subset $D\subseteq X_1\times X_2$ and maps 
	$F_i:D\to X_i$ for $i=1,2$, such that $(F_1(x,y),F_2(x,y))\subseteq D$ for every $(x,y)\in D$. 
	Let the ordered pair $(F_1,F_2)$ be such that there holds
	\begin{equation}\label{eq-cyclic-contraction}
		d(F_1(x,y),F_1(u,v))+d(F_2(z,w),F_2(t,s)\leq \alpha d(x, u)+\beta d(y,v)+\gamma d(z, t)+\delta d(w,s)
	\end{equation}
	for all $(x,y), (u,v), (z,w),(t,s)\in D$ and for some non negative constants $\alpha,\beta,\gamma,\delta$, so that $s=\max\{\alpha+\gamma,\beta+\delta\}<1$.
	Then there exists a unique pair $(\xi,\eta)$ in $D$, which is a unique coupled fixed point for the ordered pair $(F_1,F_2)$.
	Moreover the iteration sequences $\{x_{n}\}_{n=0}^\infty$ and $\{y_n\}_{n=0}^\infty$, defined in Definition \ref{iterated_sequence}
	converge to $\xi$ and $\eta$ respectively, for~any arbitrary chosen initial guess $(x,y)\in X_1\times X_2$ and error estimates hold.	
	
	If in addition $F_2(x,y)=F_1(y,x)$ then the coupled fixed point $(x,y)$ satisfies $x=y$.
\end{Theorem}

Let us consider Theorem \ref{th-HR-new} for $X_1$, $X_2$ be nonempty and closed subsets of a complete metric space $(X,d)$, rather than being
subsets of two different metric spaces with constants $\beta=\gamma=0$.
If we put $u=x$ and $v=y$ in (\ref{eq-cyclic-contraction}) we get 
\begin{equation}\label{eq-cyclic-contraction-2}
	d(F_1(x,y),F_1(u,v))+d(F_2(x,y),F_2(u,v)\leq \alpha d(x, u)+\beta d(y,v)+\gamma d(x, u)+\delta d(y,v)\leq s(d(x, u)+d(y,v)),
\end{equation}
where $s=\max\{\alpha+\gamma,\beta+\delta\}<1$.
Therefore Theorem \ref{th:3456} is a consequence of Theorem \ref{th-HR-new}.

{\bf Remark:} If in Theorem \ref{th-HR-new} $k_1=k_3=0$ we get a generalization of a Kannan type of contraction for coupled fixed points.
If in Theorem \ref{th-HR-new} $k_1=k_2=0$ we get a generalization of a Chatterjea type of contraction for coupled fixed points.

\subsection{\normalsize Application in the investigation of market equilibrium in duopoly markets}

\subsubsection{\normalsize The basic model}

Assuming we have two companies competing over the same customers \cite{Friedman} and striving to meet the demand with overall production of $Z=x+y$. The market price is defined as $P(Z)=P(x+y)$, which is the inverse of the demand function. Cost functions are $c_1(x)$ and $c_2(y)$, respectively.
Assuming that both companies are rational, the profit functions are
$\Pi_1(x,y)=xP(x+y)-c_1(x)$ and $\Pi_2(x,y)=yP(x+y)-c_2(y)$.
The goal of each player is maximizing profits, i.e. $\max\{\Pi_1(x,y):x,\ \mbox{assuming that}\ y\ \mbox{is fixed}\}$
and $\max\{\Pi_2(x,y):y,\ \mbox{assuming that}\ x\ \mbox{is fixed}\}$.
Provided that functions $P$ and $c_i$, $i=1,2$ are differentiable, we get the equations
\begin{equation}\label{equation:1a}
	\left|
	\begin{array}{l}
		\frac{\partial\Pi_1(x,y)}{\partial x}=P(x+y)+xP^\prime(x+y)-c_1^\prime(x)=0\\
		\frac{\partial\Pi_2(x,y)}{\partial y}=P(x+y)+yP^\prime(x+y)-c_2^\prime(y)=0.
	\end{array}
	\right.
\end{equation}

The solution of (\ref{equation:1a}) presents the equilibrium pair of production \cite{Friedman}.
Often equations (\ref{equation:1a}) have solutions in the form of $x=b_1(y)$ and $y=b_2(x)$, which are called response functions \cite{Friedman}.

It may turn out difficult or impossible to solve (\ref{equation:1a}) thus it is often advised to search for an approximate solution. An important drawback however is that it can be not stable.
Fortunately we can find an implicit formula for the response function in (\ref{equation:1a}) i.e.
$x=\frac{\partial\Pi_1(x,y)}{\partial x}+x=F_1(x,y) $ and $y=\frac{\partial\Pi_2(x,y)}{\partial y}+y=F_2(x,y)$.

We may end up with response functions, that do not lead to maximization of the profit $\Pi$. It is often assumed, each participant's response depends its own output as well as that of others. E.g. if at a moment $n$ the output quantities are $(x_n,y_n)$, and the first player changes its productions to $x_{n+1}=F_1(x_n,y_n)$, then the second one will also change its output to $y_{n+1}=F_2(x_n,y_n)$. 
We reach an equilibrium if there are $x$ and $y$, such that $x=F(x,y)$ and $y=f(x,y)$. To ensure that the solutions of (\ref{equation:1a}) will present a maximization of the payoff functions a sufficient condition is that $\Pi_i$ be concave or (\ref{equation:2}) is satisfied  \cite{MS}.
\begin{equation}\label{equation:2}
	\left|
	\begin{array}{l}
		\frac{\partial^2\Pi_1(x,y)}{\partial x^2}(x_0,y_0)\leq 0\\
		\frac{\partial^2\Pi_2(x,y)}{\partial y^2}(x_0,y_0)\leq 0.
	\end{array}
	\right.
\end{equation}
By using of response functions we alter the maximization problem into a coupled fixed point one thus all assumptions of concavity and differentiability can be skipped. The problem of solving the equations $x=F_1(x,y)$ and $y=F_2(x,y)$ is the problem of finding of coupled fixed points for an ordered pair of maps $(F_1,F_2)$ \cite{GL}. Yet an important limitation may be that players cannot change output too fast and thus the player may not perform maximize their profits.

\subsubsection{Connection between the second order conditions and the contraction type conditions}

We will restate Theorem \ref{th-HR-new} for $k_2=k_3=0$, $X_1, X_2$ be subsets of a metric space $(X,d)$ in the economic language. 

\begin{Assumption}\label{assumption1x} Let there is a duopoly market, satisfying the following assumptions:
	\begin{enumerate}\itemsep=0pt
		
		\item The two firms are producing homogeneous goods that are perfect substitutes.
		
		\item The first firm can produce qualities from the set $X_1$ and the second firm can produce qualities from the set $X_2$, where $X_1$ and $X_2$ be closed, 
		nonempty subsets of a complete metric space $(X,d)$
		
		\item Let there exist a closed subset $D\subseteq X_1\times X_2$ and maps $F_i:D\to X_i$, such that $(F_1(x,y),F_2(x,y))\subseteq D$ for every $(x,y)\in D$, 
		be the response functions for firm one and two respectively
		
		\item Let there exist $\alpha <1$, such that the inequality
		\begin{equation}\label{equation:2xx}
			d(F_1(x,y),F_1(u,v))+d(F_2(x,y),F_2(u,v))\leq\alpha (d(x, u)+\rho(y,v))
		\end{equation}
		holds for all $(x,y), (u,v)\in X_1\times X_2$.
	\end{enumerate}
\end{Assumption}

Then there exists a unique pair $(\xi,\eta)$ in $D$, such that $\xi=F_1(\xi,\eta)$ and $\eta=F_2(\xi,\eta)$,
i.e. a market equilibrium pair.

If in addition $F_2(x,y)=F_1(y,x)$ then the coupled fixed point $(\xi,\eta)$ satisfies $\xi=\eta$.

{\bf Example 1:}
Let us we get the response functions  $F_i$ through the maximization of the payoff functions 
$\Pi_i(x,y)$. Let $(x_0,y_0)$ be a unique solution of (\ref{equation:1a}), then it is well known that the optimization of the payoff functions in Cournout model is guaranteed if (\ref{equation:2}) is satisfied.

We will show that from (\ref{equation:2xx}) follows (\ref{equation:2}). 

Following \cite{DKRZ} the response functions are defined as
\begin{equation}\label{equation:4x}
	F_1(x,y)=\textstyle\frac{\partial\Pi_1(x,y)}{\partial x}+x\ \ \mbox{and}\ \
	F_2(x,y)=\frac{\partial\Pi_2(x,y)}{\partial y}+y.
\end{equation}
From (\ref{equation:2xx}) we get
\begin{equation}\label{limit1}
	\lim_{\Delta x\to 0}\textstyle\frac{|F_1(x+\Delta x,y)-F_1(x,y)|+|F_2(x+\Delta x,y)-F_2(x,y)|}{\Delta x}\leq\alpha \ \
	\lim_{\Delta x\to 0}\textstyle\frac{|F_1(x,y+\Delta y)-F_1(x,y)|+|F_2(x,y+\Delta y)-F_2(x,y)|}{\Delta y}\leq\alpha
\end{equation}
and therefore it follows that $\left|\frac{\partial F_1}{\partial x}(x_0,y_0)\right|\leq\alpha<1$ and $\left|\frac{\partial F_2}{\partial y}(x_0,y_0)\right|\leq\alpha <1$.
Then from (\ref{equation:4x}) we get
$$
\frac{\partial^2\Pi_1(x,y)}{\partial x^2}(x_0,y_0)=\frac{\partial F_1}{\partial x}(x_0,y_0)-1<|\alpha| -1<0
$$
and
$$
\frac{\partial^2\Pi_2(x,y)}{\partial y^2}(x_0,y_0)=\frac{\partial F_2}{\partial y}(x_0,y_0)-1<|\alpha| -1<0.
$$

A similar condition to (\ref{equation:2x}) is investigated in \cite{KabZla}, where maps with the mixed monotone property are considered.
In this case (\ref{equation:2xx}) holds only for part of the variables and therefore we can not 
take limits in (\ref{limit1}). Thus the response functions from \cite{KabZla} may be not differentiable.

Besides presenting of sufficient conditions for the existence of a market equilibrium Assumption \ref{assumption1x} gives
also sufficient conditions for the stability of the process of the consecutive responses of the player in they do not change their behavior. 

{\bf Example 2}
Let us consider a model with $P(x,y)=100-x-y$ and cost functions $C_1(x)=\frac{x^2}{2}$ and $C_2(y)=\frac{y^2}{2}$. By (\ref{equation:1a}) we get 
\begin{equation}\label{equation:23}
	\left|
	\begin{array}{l}
		\frac{\partial\Pi_1(x,y)}{\partial x}=100-3x-y=0\\
		\frac{\partial\Pi_2(x,y)}{\partial y}=100-x-3y=0.
	\end{array}
	\right.
\end{equation}
the second order conditions are $\frac{\partial^2\Pi_1(x,y)}{\partial x^2}=-3<0$ and
$\frac{\partial^2\Pi_2(x,y)}{\partial y^2}=-3<0$ and consequently the solution of the system of equations (\ref{equation:23}) is the equilibrium points, because it satisfies (\ref{equation:2}). Unfortunately the response functions in the model will be $F(x,y)=100-2x-y$ and $f(x,y)=100-x-2y$, which will not satisfy condition (\ref{equation:2xx}).

\begin{table}[H]%
\caption{Values of the iterated sequence $(x_n,y_n)$ if stared with $(20,30)$.}\label{tbl2}
\centering %
\begin{tabular}{ccccccc}
$n$&0&1&2&$\dots$&2k&2k+1\\
\hline\\
$x_n$&20&30&20&$\dots$&20&30\\
$y_n$&30&20&30&$\dots$&30&20\\
\end {tabular}
\end {table}

If the initial start is different we get Table \ref{tbl3}.

\begin{table}[H]%
	\caption{Values of the iterated sequence $(x_n,y_n)$ if stared with $(20,31)$.}\label{tbl3}
	\centering %
	\begin{tabular}{cccccccc}
		$n$&0&1&2&3&4&5&6\\
		\hline\\
		$x_n$&20&29&24&17&60&0&100\\
		$y_n$&31&18&35&6&71&0&100\\
		\end {tabular}
		\end {table}

In both cases Table \ref{tbl2} and Table \ref{tbl3} we see that the process is not converging.

Let us point out that the system (\ref{equation:1a}) may have more than one solution $(x,y)$, satisfying the second order conditions
(\ref{equation:2}). In this case we will need further investigation to find which one of the solutions is the solution of the optimization problem
of Cournot's model. Therefore never mind that (\ref{equation:2xx})  is a stronger restriction than (\ref{equation:2})
the model from Assumption \ref{assumption1x} is a different from the well known Cournot's optimization problem.

\subsubsection{Comments on the coefficients $\alpha$, $\beta$, $\gamma$ and $\delta$ in Theorem \ref{th:3456}}

Although the Theorem \ref{th:3456} is a consequence of Theorem \ref{th-HR-new} it seems that the usage of four coefficients
may give better understanding of duopoly markets. 

Let the response functions $F_1$ and $F_2$ satisfy
\begin{equation}\label{equation:21}
	\rho(F_1(x,y),F_1(u,v))\leq \alpha \rho(x, u)+\beta \rho(y,v)
\end{equation}
and
\begin{equation}\label{equation:22}
	\rho(F_2(x,y),F_2(u,v))\leq \gamma\rho(x, u)+\delta \rho(y,v).
\end{equation}
If $\max\{\alpha+\gamma,\beta+\delta\}\in (0,1)$, then by summing up (\ref{equation:21}) and (\ref{equation:22})  the model satisfies inequality (\ref{eq-cyclic-contraction}). 
Let us assume that $\alpha$ and $\delta$ are close to $1$ and $\beta$ and $\gamma$ are close to $0$. This means that both players do not pay too much attention to the behavior of the production of the other one. they are interested mostly of their productions.

{\bf Example 3}
Let us consider a model with the following response functions $F(x,y)=45-0.98x-0.09y$ and $f(x,y)=50-0.01x-0.9y$. An example of Cournot model can be considered
$P(x,y)=50-0.09x-0.01y$, and cost functions $C_1(x)=0.985x^2$ and $C_2=0.86y^2$.
Thus we get that
$$
|x_{n+2}-x_{n+1}|=|F_1(x_n,y_n)-F_1(x_{n+1},y_{n+1})|\leq 0.98|x_n-x_{n+1}|+0.09|y_n-y_{n+1}|
$$
and
$$
|y_{n+2}-y_{n+1}|=|F_2(x_n,y_n)-F_2(x_{n+1},y_{n+1})|\leq 0.01|x_n-x_{n+1}|+0.9|y_n-y_{n+1}|,
$$
which can be interpenetrated as any player take account just on his change of the production.
The market equilibrium is $(24.06,26.18)$

\begin{table}[H]%
	\caption{Values of the iterated sequence $(x_n,y_n)$ if stared with $(10,30)$.}\label{tbl1}
	\centering %
	\begin{tabular}{ccccccccccccccc}
		$n$&0&1&2&3&4&5&10&21&50&51&120&121&599&600\\
		\hline
		$x_n$&10&37&12&35&13&33.7&16.8&30.8&21.1&26.9&22.64&25.43&24.07&24.05\\
		$y_n$&30&18&33&20&31&21.4&28.6&24.1&25.8&26.4&26.03&26.34&26.19&26.18\\
		\end {tabular}
		\end {table}

We see from the Table \ref{tbl1} that at the very beginning the osculations of the sequence of productions are big and it take a lot of time to get close enough to the equilibrium values.

\subsubsection{Some applications on newly investigated oligopoly models}

A deep analysis of a class of oligopoly markets is presented in \cite{AEJ}. In section $2$ in \cite{AEJ} authors analyze market equilibrium, obtained by the use of the first and second order conditions. They have assumed $P(Q)=Q^{-1/\eta}$, where $P$ be the market price, 
$x,y\geq 0$ are the quantity supplied by firm one and two, respectively, $Q=x+y$ be the total output and $\eta>0$ be a parameter. Both players 
share a linear cost function with constant average and marginal cost $c_i>0$. 
As far as part of the results in \cite{AEJ} are for $c_i=c$ for $i=1,2$, let us assume that $c_1=c_2=c$.
The first order conditions in \cite{AEJ} yield to the system of equations:
$$
\left|\begin{array}{l}
	x=\eta Q-c\eta Q^{1+\frac{1}{\eta}}\\
	y=\eta Q-c\eta Q^{1+\frac{1}{\eta}}.
\end{array}\right.
$$
Both players share one and the same response function
$F_1(x,y)=F_2(x,y)=F(x,y)=\eta Q-c\eta Q^{1+\frac{1}{\eta}}$, where $Q=x+y$. Using the mean value theorem we get that there holds the equality
$$
|F(x,y)-F(u,v)|=\left|(\eta-c(1+\eta)Q_\lambda^{1/\eta})\right|(|x-u|+|y-v|)\leq(\eta+c(1+\eta)Q_\lambda^{1/\eta})(|x-u|+|y-v|)
$$
where $Q_{\lambda}=\lambda (x+y)+(1-\lambda)(u+v)$ for some $\lambda\in (0,1)$.
As far as the total output of the economy is bounded from above we can assume that
$Q_{\lambda}\leq Q_{\max}<+\infty$ we get
$$
|F_1(x,y)-F_1(u,v)|+|F_2(x,y)-F_2(u,v)|\leq 2\left|(\eta+c(1+\eta)Q_{\max}^{1/\eta})\right|(|x-u|+|y-v|).
$$

Assumption \ref{assumption1x} can be applied if
$2\left|\eta+c(1+\eta)Q^{1/\eta}_{\max}\right|<1$, i.e
\begin{equation}\label{ww1}
	0<cQ^{1/\eta}_{\max}< \frac{1-2\eta}{2(1+\eta)}<1,
\end{equation} 
which can holds true if $\eta\in [0,1/2)$.

The analysis in \cite{AEJ}, using of the second order conditions yields that
there exists a market equilibrium if $\eta\geq \max\left\{1, \frac{x}{2y-x},\frac{y}{2x-y}\right\}$. If $\eta<1$, following \cite{AEJ}, then the second order conditions
do not ensure an existence of a market equilibrium. 
It follows that whenever the marginal costs $c$ satisfies (\ref{ww1}) there exists a unique market equilibrium.
Thus Assumption \ref{assumption1x} covers and cases that are not covered by the classical 
first and second order conditions.

\subsubsection{A generalized response function}

When considering a real life model of duopolies we need to pay attention to the surplus of the total production.
Actually every one of the participants in the market takes into account not only the the productions, that were realized
on the market, but also he pays attention of his surplus quantities. Let us denote the set of the possible productions of 
player $i$ by $U_i$; the set of the realized production on the market by $P_i\subseteq U_i$; the set of its surplus quantities
by $s_i$, $i=1,2$. Let us put $X_i=P_i\times s_i$. Each of the players is not able to know the surplus production of the other one. 
Therefore a more real model of the response functions of the two player will be
$$
f_1:X_1\times P_2\to U_1,\ \ f_2:X_2\times P_1\to U_2.
$$
Starting at a moment $t_0$ with realized on the market productions $p^{(0)}_i$, surpluses $s^{(0)}_i$ and productions $u^{(0)}_i$ , $i=1,2$ for both players
it results to a new productions of the players 
$$
u^{(1)}_1=f_1\left(p^{(0)}_1,p^{(0)}_2,s^{(0)}_1\right)\in U_i,\ \ u^{(1)}_2=f_2\left(p^{(0)}_1,p^{(0)}_2,s^{(0)}_2\right)\in U_2.
$$
The market reacts to this new levels of production by generating new surplus quantities 
$s^{(1)}_i=Q_i(u^{(1)}_1,u^{(1)}_2)$, where $Q_i:U_1\times U_2\to U_i$, $i=1,2$ be the responses of the
market to the produced quantities of both players. Thus the realized quantities on the market for each of the players at moment $t_1$ will be
$$
\begin{array}{lll}
	p^{(1)}_1&=&u^{(1)}_1-s^{(1)}_1=f_1\left(p^{(0)}_1,p^{(0)}_2,s^{(0)}_1\right)-Q_1\left(u^{(1)}_1,u^{(1)}_2\right)\\
	&=&f_1\left(p^{(0)}_1,p^{(0)}_2,s^{(0)}_1\right)-Q_1\left(f_1\left(p^{(0)}_1,p^{(0)}_2,s^{(0)}_1\right),f_2\left(p^{(0)}_1,p^{(0)}_2,s^{(0)}_2\right)\right)
\end{array}
$$
and
$$
\begin{array}{lll}
	p^{(1)}_2&=&u^{(1)}_2-s^{(1)}_2=f_2(p^{(0)}_1,p^{(0)}_2,s^{(0)}_2)-Q_2(u^{(1)}_1,u^{(1)}_2)\\
	&=&f_2(p^{(0)}_1,p^{(0)}_2,s^{(0)}_2)-Q_2(f_1(p^{(0)}_1,p^{(0)}_2,s^{(0)}_1),f_2(p^{(0)}_1,p^{(0)}_2,s^{(0)}_2)).
\end{array}
$$
We will define a new function, which we will call a generalized response function of the player and the market.
Let $X\in X_1$, $Y\in X_2$, i.e. $X=(x,\delta x)\in P_1\times s_1$ and $Y=(y,\delta y)\in P_2\times s_2$.
$$
F_1(X,Y)=F_1(x,y,\delta x,\delta y)=(f_1(x,y,\delta x)-Q_1(f_1(x,y,\delta x),f_2(x,y,\delta y)),Q_1(f_1(x,y,\delta x),f_2(x,y,\delta y)))
$$
and
$$
F_2(X,Y)=F_2(x,y,\delta x,\delta y)=(f_2(x,y,\delta y)-Q_2(f_1(x,y,\delta x),f_2(x,y,\delta y)),Q_2(f_1(x,y,\delta x),f_2(x,y,\delta y))).
$$
As far as in Assumption \ref{assumption1x} the sets $X$ and $Y$ can be subsets of $\mathbb{R}^n$ we can reformulate
Assumption \ref{assumption1x} for the case of the generalized response function of the player and the market:
\begin{Assumption}\label{assumption11}Let there is a duopoly market, satisfying the following assumptions:
	\begin{enumerate}\itemsep=0pt 
		\item The two firms are producing homogeneous goods that are perfect substitutes.
		\item The firm $i$, $i=1,2$ can produce qualities from the set $U_i$, its set of the realized on the market production be $P_i$ and 
		the set of its surplus productions be $s_i$, where $X=P_1\times s_1$ and $Y=P_2\times s_2$ be closed, nonempty subsets of a complete metric space $(Z,\rho)$
		\item Let there exist a closed subset $D\subseteq X\times Y$ and maps $F_1:D\to X$ and $F_2:D\to Y$, such that 
		$(F_1(x,y),F_2(x,y))\subseteq D$ for every $(x,y)\in D$, be the generalized response function of the player and the market for firm one and two respectively
		\item Let there exists $\alpha\in (0,1)$, such that the inequality
		\begin{equation}\label{equation:22-b}
			\rho(F_1(x,y),F_1(u,v))+\rho(F_2(x,y),F_2(u,v))\leq\alpha (\rho(x, u)+\rho(y,v))
		\end{equation}
		holds for all $(x,y), (u,v)\in X\times Y$.
	\end{enumerate}
\end{Assumption}

We will illustrate Assumption \ref{assumption11} by an example.
Let $U_i=[0,+\infty)$, $P_i=[0,+\infty)$, $s_i=[0,+\infty)$, $X=P_1\times s_1$ and $Y=P_2\times s_2$.
Let $X$ and $Y$ be subsets of $(\mathbb{R}^2,\rho_1)$, where $d_1((x_1,y_1),(x_2,y_2))=|x_1-x_2|+|y_1-y_2|$. 
Let $(X\times Y)$ be endowed with the metric $\rho(\cdot,\cdot)=d_1(\cdot,\cdot)+d_1(\cdot,\cdot)$.
Let $f_1:X\times P_2\to U_1$ and $f_2:Y\times P_1\to U_2$ be defined by
$f_1(x,y,\delta x)=45-0.5x+0.25y-0.1\delta x$ and $f_2(x,y,\delta y)=20-0.2x-0.25y-0.05\delta y$,
where $(x,\delta x)\in X$ and $(y,\delta y)\in Y$.
Let the response functions of the market $Q_1:U_1\times U_2\to U_1$ and $Q_2:U_1\times U_2\to U_2$
be defined by $Q_1(x,y)=0.05x+0.03y$ and $Q_2(x,y)=0.04x+0.06y$.
Let the generalized response function of the player and the market $F_1:X\times Y\to X$ and $F_2:X\times Y\to Y$ be
$$
F_1(x,y,\delta x,\delta y)=(f_1(x,y,\delta x)-Q_1(f_1(x,y,\delta x),f_2(x,y,\delta y)),Q_1(f_1(x,y,\delta x),f_2(x,y,\delta y)))
$$
and 
$$
F_2(x,y,\delta x,\delta y)=(f_2(x,y,\delta y)-Q_2(f_1(x,y,\delta x),f_2(x,y,\delta y)),Q_2(f_1(x,y,\delta x),f_2(x,y,\delta y))).
$$

We need to show only that (\ref{equation:22-b}) holds true.

From
$$
\begin{array}{lll}
	\rho_1(F_1(x,y,\delta x,\delta y),F_1(u,v,\delta u,\delta v))&\leq&
	0.5|x-u|+0.1|\delta x -\delta u|+0.25|y-v|+0.3|\delta y-\delta v|\\
	&\leq&0.5\rho_1((x,y,\delta x,\delta y),(u,v,\delta u,\delta v)).
\end{array}
$$
and
$$
\begin{array}{lll}
	\rho_1(F_2(x,y,\delta x,\delta y),F_2(u,v,\delta u,\delta v))&\leq&
	0.2|x-u|+0.5|\delta x -\delta u|+0.25|y-v|+0.1|\delta y -\delta v|\\
	&\leq&0.5\rho_1((x,y,\delta x,\delta y),(u,v,\delta u,\delta v)).
\end{array}
$$
Thus (\ref{equation:22-b}) is satisfied with $\alpha\leq 0.5$.

The equilibrium solution of the market are $x=27.1$, $y=9.6$, $\delta x=1.6$ and $\delta y=1.2$. The example show that in the equilibrium both players will have surplus productions greater that zero.

If we suppose that the players do not pay attention to the surplus quantities, i.e  
$F_1(x,y,\delta x)=45-0.5x+0.25y$ and $F_2(x,y,\delta y)=20-0.2x-0.25y$,
we get an equilibrium solution in the market $x=29.8$ and $y=11.2$.

\subsubsection{Applications of Theorem \ref{th-HR-new} for optimization of non--differentiable payoff functions and examples}

It seems from Theorem \ref{th:3456} that we can impose different type of contraction conditions that will be not equivalent to 
(\ref{equation:2}). Let us restate Theorem \ref{th-HR-new}, when $k_1=k_2=0$ in the economic language.

\begin{Assumption}\label{assumption1} Let there is a duopoly market, satisfying the following assumptions:
	\begin{enumerate}\itemsep=0pt
		
		\item The two firms are producing homogeneous goods that are perfect substitutes.
		
		\item The first firm can produce qualities from the set $X_1$ and the second firm can produce qualities from the set $X_2$, where $X_1$ and $X_2$ be closed, 
		nonempty subsets of a complete metric space $(X,d)$
		
		\item Let there exist a closed subset $D\subseteq X_1\times X_2$ and maps $F_i:D\to X_i$, such that $(F_1(x,y),F_2(x,y))\subseteq D$ for every $(x,y)\in D$, 
		be the response functions for firm one and two respectively
		
		\item Let there exist $\beta\in [0,1/2)$, such that the inequality
		\begin{equation}\label{equation:2x}
			\sum_{i=1}^2 d(F_i(x,y),F_i(u,v))\leq \beta(d(x,F_1(x,y))+d(y,F_2(x,y))+d(u,F_1(u,v))+d(v,F_2(u,v)))
		\end{equation}
		holds for all $(x,y), (u,v)\in D$.
	\end{enumerate}
\end{Assumption}

Then there exists a unique pair $(\xi,\eta)$ in $D$, such that $\xi=F(\xi,\eta)$ and $\eta=f(\xi,\eta)$,
i.e.,~a~market equilibrium pair.
Moreover the iteration sequences $\{x_{n}\}_{n=0}^\infty$ and $\{y_n\}_{n=0}^\infty$, defined in Definition \ref{iterated_sequence}
converge to $\xi$ and $\eta$ respectively and the error estimates hold.

If in addition $F_2(x.y)=F_1(y,x)$ then the coupled fixed point $(x,y)$ satisfies $x=y$.

{\bf Example 4:}
Let us consider a market with two competing firms, producing perfect substitute products.
Let us consider the response functions of player one and two be
$$
F_1(x)=F_1(x,y)=\left\{
\begin{array}{rl}
	0.2&x\in [0,0.8]\\
	0.1&x\in (0.8,1]
\end{array}
\right.\  \ 
F_2(y)=F_2(x,y)=\left\{
\begin{array}{rl}
	0.9&y\in [0,0.1]\\
	0.8&y\in (0.1,1],
\end{array}
\right.
$$
respectively. We can choose $D$ to be $D=[0,1]\times [0,1]$. 
It is easy to check that $F_1:D\to [0,1]$, $F_2:D\to [0,1]$ and $(F_1(D),F_2(D))\subseteq D$.

We will consider several cases to show that $(F_1,F_2)$ satisfies (\ref{equation:2x}).

Let $x,u\in [0,0.8]$ or $x,u\in (0.8,1]$ then
$\left|F(x)-F(u)\right|=0\leq\beta_1 |x-F(x)|+\beta_1 |u-F(u)|$
holds for any $\beta_1\in [0,1/2)$.

\begin{figure}[h!]
	\centering
	\includegraphics[width=0.35\textwidth]{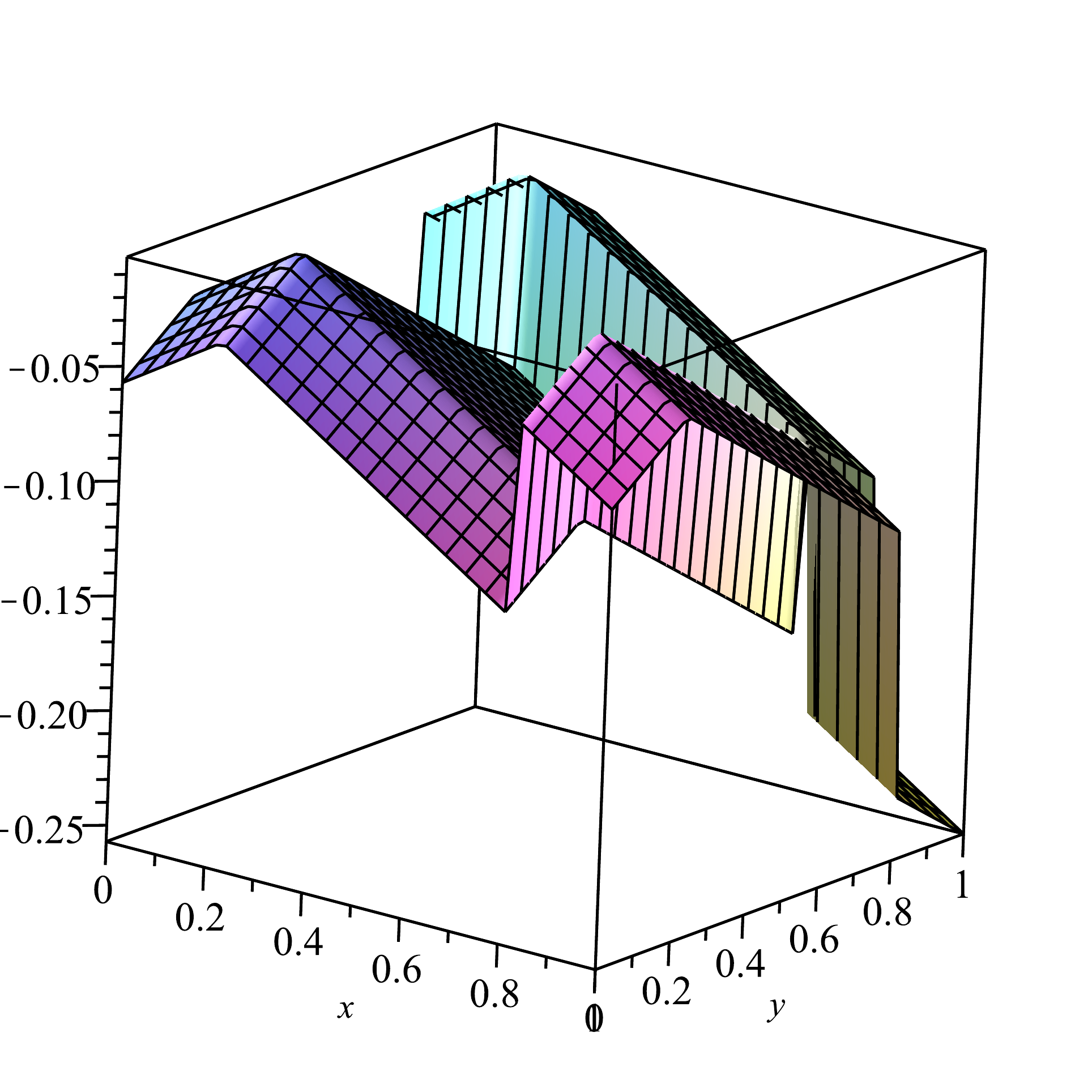}
	\caption{Graphic of the function $|F(x)-F(y)|-\frac{1}{7}(|x-F(x)|+|y-F(y)|)$.}
	\label{fig1-1}
\end{figure}

Let $x,\in [0,0.8]$ and $u\in (0.8,1]$ then, using the equalities $0=\inf\{|x-F(x)|:x\in [0,0.8]\}$
and $0.7=\inf\{|u-F(u)|:u\in (0.8,1]\}$, we get that the inequality
$\left|F(x)-F(u)\right|=0.1\leq\beta_1 0+\beta_1 0.7\leq\beta_1 |x-F(x)|+\beta_1 |u-F(u)|$
will hold for any $\beta_1\in [1/7,1/2)$ (Fig. \ref{fig1-1}).

Let $y,v\in [0,0.1]$ or $y,v\in (0.1,1]$ then
$\left|F_2(y)-F_2(v)\right|=0\leq\beta_2 |y-F_2(y)|+\beta_2 |v-F_2(v)|$
holds for any $\beta_2\in [0,1/2)$.

\begin{figure}[h!]
	\centering
	\includegraphics[width=0.35\textwidth]{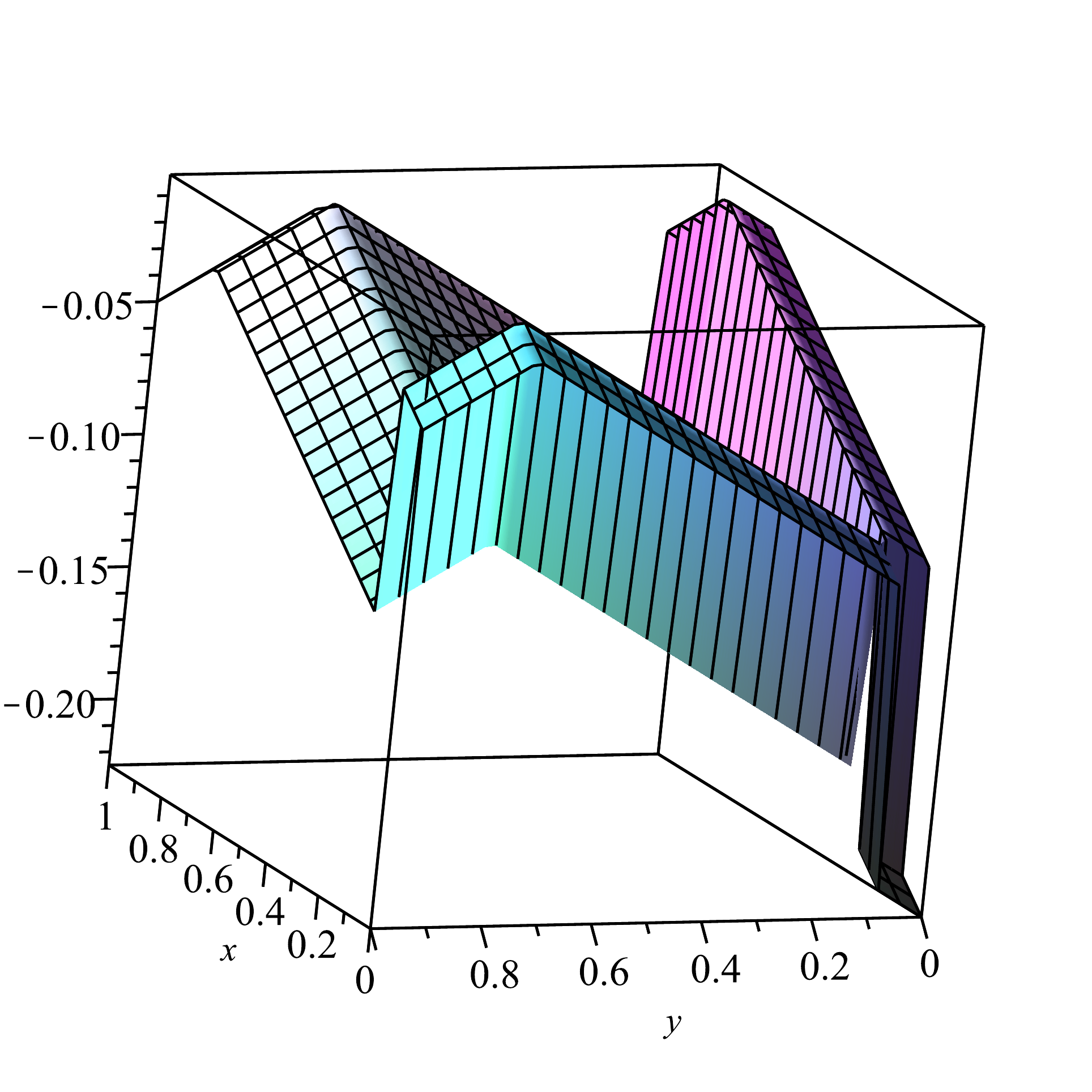}
	\caption{Graphic of the function $|F_2(y)-F_2(v)|-\frac{1}{8}(|y-F_2(y)|+|v-F_2(v)|)$.}
	\label{fig1-2}
\end{figure}

Let $y,\in [0,0.1]$ and $v\in (0.1,1]$ then, using the equalities $0.8=\inf\{|y-F_2(y)|:x\in [0,0.1]\}$
and $0=\inf\{|v-F_2(v)|:u\in (0.1,1]\}$, we get that the inequality
$\left|F_2(y)-F_2(v)\right|=0.1\leq\beta_2 0.8+\beta_2 0\leq\beta_2 |y-F_2(y)|+\beta_2 |v-F_2(v)|$
will hold for any $\beta_2\in [1/8,1/2)$ (Fig. \ref{fig1-2}).

Therefore 
$\left|F_1(x)-F_1(u)\right|+\left|F_2(y)-F_2(v)\right|\leq 
\displaystyle\frac{1}{7}\left(|x-F_1(x)|+|u-F_1(u)|+|y-F_2(y)|+|v-F_2(v)|\right)$
and thus  the ordered pair $(F_1,F_2)$ satisfies Assumption \ref{assumption1} with a constant $\beta=1/7$.
Consequently there exists an equilibrium pair $(x,y)$ and for any initial start in the economy the iterated sequences $(x_n,y_n)$ converge to the 
market equilibrium $(x,y)$. We get in this case that the equilibrium pair of the production of the two firms is $(0.8, 0.1)$.

The considered model with response functions $F_1$ and $F_2$ does not satisfies (\ref{equation:2xx}). Indeed let us consider 
$x=0.8$, $u=x+\varepsilon$, and $y=0.1$, $v=y+\varepsilon$.
Then
$\left|F_1(x)-F_1(u)\right|+\left|F_2(y)-F_2(v)\right|=0.1+0.1=0.2\geq 2\varepsilon=2(|x-u|+|y-v|)$ for any $\varepsilon\leq 0.1$
and thus we can not apply Assumption \ref{assumption1x}.

The example shows that if $F_i$ were obtained by solving the optimization of the payoff functions, then we can not 
speak about the second order conditions as far as $F_i$ are not differentiable.

\bibliographystyle{plain}
\bibliography{refs}
\end{document}